\documentstyle[preprint,aps]{revtex}
\def \util{\tilde{U} }
\def \rtil{\tilde{R} }
\def \mbar{\bar{\mu}}
\begin{document}
%\baselineskip1.6\baselineskip
%\parindent1.0\parindent
\title{Finite Density Effect in the Gross-Neveu 
Model in a Weakly Curved $R^1\times S^2$ Spacetime  } 
\author{Dae Kwan Kim\footnote{E-mail:dkkim@phya.yonsei.ac.kr}}
\address{ Department of Physics and Institute for Mathematical Sciences, 
Yonsei University, Seoul 120-749, Korea}
\author{K. G. Klimenko\footnote{E-mail:kklim@mx.ihep.su}}
\address{ Institute for High Energy Physics, 142284, Protvino, Moscow region, Russia}
\maketitle
\vspace{4cm}
\begin{abstract}
The three-dimensional Gross-Neveu model in $R^{1} \times S^{2}$ spacetime
is considered at finite particles number density.
We evaluate an effective potential of the composite scalar field $\sigma(x)$,
which is expressed in terms of a scalar curvature  $R$ and nonzero chemical
potential $\mu$.  We then derive the critical values of
$(R,\mu)$ at which the system undergoes the first order phase
transition from the phase with broken
chiral invariance to the symmetric phase.
\end{abstract}
\draft
%\newpage
\section{ Introduction }
Last time four-fermion field theories in $(2+1)$-dimensional
Minkowski spacetime, which are known as Gross -- Neveu (GN) models
\cite{1}, are under extensive investigation for purely
theoretical motivation and also due to their applications to planar
condensed matter physics. Such theories possess many desirable
properties: the renormalizability in $1/N$ expansion, dynamical
breaking of chiral symmetry and generation of fermion mass for a
large coupling constant as in QCD \cite{2}, the analogy to the
BCS theory of superconductivity in two spatial dimensions and
the possibility to describe the new phenomenon of high
temperature superconductivity \cite{3}, the reduction to the
$S=1/2$ quantum antiferromagnet Heisenberg model in the
continuum limit \cite{4} and so on. Main features of these
models, obtained in large $N$ expansion technique, are confirmed
within the framework of other nonperturbative approarches \cite{20}.
Since there are no closed physical systems in nature, the influence
of different external factors on the vacuum of the simplest GN
model was considered.  In \cite{5} some critical phenomena of
this theory were studied at nonzero temperature $T$ and chemical
potential $\mu$. Recently, on the same foundation a new property
of external (cromo-)magnetic fild $H$ to promote the dynamical
chiral symmetry breaking was discovered \cite{6}. (At present
it is the well known effect of dynamical chiral symmetry breaking
catalyst by external magnetic field \cite{7}, which is 
under intensive consideration \cite{8}.) The role of $T,~H$ as
well as of $\mu,~H$ in the formation of a ground state of the
GN model was also clarified \cite{9}.
The study of dynamical symmetry breaking in spacetimes with curvature
and nontrivial topology is also of great importance, since in
the early universe the gravity was sufficiently strong and one should
take it into account. There is a reach literature on this
subject (see the review \cite{10}).  The papers in \cite{41,42}
are the first ones where the effect of curvature and nontrivial
topology on the chiral symmetry breaking in four - fermion
models was discussed.  The curvature - induced first order phase
transition from a chiral symmetric to a chiral nonsymmetric
phase was shown to exist in that models in the linear curvature
approximation. It turns out that in specific spacetimes such as
Einstein universe \cite{43} and maximally symmetric spacetimes
\cite{44} the above mentioned models may be solved exactly in the
leading order of large $N$ expansion technique.  Finally,
dynamical symmetry breaking in the external gravitational and
magnetic fields was considered \cite{45}.
It is well - known that low dimensional four - fermion field
theories, especially the (2+1) - dimensional GN model,
in curved spacetimes \cite{42,43,11,12} and in the nonsimply connected 
spacetimes \cite{30,13} may be very useful for the
investigation of physical processes in thin films and in the
materials with layer structure. The matter is that external
stress, applied to the planar system, may change topology and curvature
of a surface. The consideration of above mentioned low dimensional models
in curved spaces is an interesting subject by itself,
since in these cases the quantum theories can be computed exactly.
The great amount of observable physical phenomena are due to 
nonzero particle density (superconductivity, quantum Hall effect etc.).
So in the present paper the influence of both chemical potential and
curvature of space on the phase structure of (2+1) - dimensional GN model
is studied. Especially, we shall consider $R^1\times S^2$ spacetime to
clarify our discussion (the effect of temperature on the vacuum of GN model
in this spacetime was considered in \cite{10}).        
In Section II, we evaluate
the one-loop effective potential in $R^{1} \times S^{2}$
spacetime at nonzero chemical potential. Presented in Section III is the
detailed analysis of the effective potential, which shows the existance
of a phase transition restoring the chiral symmetry 
of the system as the
curvature $R$ and chemical potential $\mu$ are varied.
 Finally, we summarize our results in Section IV.
\section{  Effective potential in $R^{1} \times S^{2}$ spacetime at $\mu\neq 0$}
The $R^{1} \times S^{2}$ space-time is chosen to be 
\begin{equation}
ds^{2}= dt^{2}- a^{2}( d \theta^{2} +\sin^{2} \theta \; d \phi^{2}), 
\end{equation}
where $a$ is its constant radius.
The four-fermion model in this spacetime is described by the 
action \cite{10,43} 
\begin{equation}
S= \int d^{3} x \sqrt{-g} \left[i \bar{\psi}_{j}  \gamma^{\mu} (x) 
\nabla_{\mu} \psi_{j} + \frac{ \lambda^{2}}{2 N} ( \bar{\psi}_{j} \psi_{j} )^{2}  \right] ,
\end{equation}
where $ \nabla_{\mu}$ is the covariant derivative and the
summation over $j$ is implied ($j=1,2,..,N$). Here fermion
fields $\psi_j$ are taken in the reducible four dimensional
representation of $SL(2,C)$. For this case the algebra of the
$\gamma$ - matrices is presented
in \cite{2}.
This action has the discrete chiral symmetry, 
\begin{equation} 
\psi \rightarrow  \gamma_{5} \psi  . 
\end{equation} 
As a result, the chiral symmetry is maintained at any
order of ordinary perturbation theory. However as can be seen in different
nonperturbative approaches \cite{1,2,20}
the symmetry may be broken dynamically for large values of
coupling constant $\lambda$.
To see the nonperturbative features such as spontaneous symmetry breaking
and dynamical mass generation in the present model, it is convenient
to rewrite above action in an equivalent form \cite{1} by introducing the 
auxiliary field $\sigma(x)$,
\begin{equation}
 S= \int d^{3} x \sqrt{-g} \left[i \bar{\psi_{j}}  \gamma^{\mu} (x) \nabla_{\mu} \psi_{j} -  \sigma  \bar{\psi_{j}} \psi_{j} - \frac{N}
{2 \lambda^{2} } \sigma^{2} \right] .
\end{equation} 
This expression explicitly says that the vacuum expectation
value of $\sigma$ field plays the role of mass for the
fermions.
In order to find the effective potential in the theory with the
action Eq. (2) we follow \cite{41,42} where this quantity 
was considered in a weak curvature approximation.  First of all
let us integrate over the fermion fields in Eq. (4) and evaluate
an effective action $ S_{eff}( \sigma)$ describing the
self-interaction of $\sigma$ field:
\begin{equation}
\exp(i N S_{eff}( \sigma) ) = \int D\psi D\bar{\psi} \exp[ i S( \psi, \bar{\psi}, \sigma )].
\end{equation} 
Here we use the $1/N$ expansion which is the fermion-loop 
expansion. 
In the mean-field approximation, where the $\sigma (x)$ field is assumed to 
be constant, and to the leading order in the large $N$, one can obtain
 the one-loop effective potential $U(\sigma)$ from the action
$S_{eff}(\sigma)$:
\begin{equation}
U(\sigma) = \frac{ \sigma^{2}}{2 \lambda^{2}}  
+i \mbox{ tr} \langle x |\;\ln ( i \gamma^{\mu} (x) \nabla_{\mu}-  \sigma ) \; | x \rangle,  
\end{equation}
where tr is over indices other than spacetime indices.
Using the Green function $G_{F}(x,y;\sigma)$ defined by the relation
\begin{equation}
G_{F}(x,y;\sigma ) \equiv \langle x | (i \gamma^{\mu}  \nabla_{\mu}-  \sigma)^{-1} | y \rangle,  
\end{equation}
we rewrite Eq. (6) as follows:
\begin{equation}
U(\sigma) = \frac{ \sigma^{2}}{2 \lambda^{2}}  
-i \mbox{ tr} \ln G_{F}(x,x; \sigma).
\end{equation}
The logarithm  may be eliminated from this equation by introducing
the parameter $s$:
\begin{equation}
\ln \left[ \frac{ K-\sigma}{K } \right] = -\int_{0}^{\sigma} d s \frac{1}{ K -s},
\end{equation}
where an operator $K$ is given as
 $i \gamma^{\mu} (x) \nabla _{\mu}$ in the present case.  
Therefore, Eq. (8) is rewritten in the following form:
\begin{equation}
U(\sigma)= \frac{\sigma^{2}}{2 \lambda^{2} } -i \mbox{ tr} \int_{0}^{ \sigma} d s \int 
\frac{d^{3} k }{(2 \pi)^{3}}\; G_{F}(k;s),
\end{equation}
where the momentum-space Green function $G_{F}(k;s)$ has been used.
Now one may introduce the Riemann normal coordinate \cite{14}
with origin at any point in the spacetime. In this coordinate
system we use the following expression for the Green function $G_{F}(k;s)$:
\begin{eqnarray}
G_{F}(k;s)  &=&   \frac{\gamma^{a} k_{a} +s}{ 
 k^{2} -s^{2}} -
 \frac{R}{12}  \frac{\gamma^{a} k_{a} +s} { (k^{2} -s^{2}) ^{2} } 
        + \frac{2}{3} R_{\mu \nu} k^{\mu} k^{\nu}  \frac{(\gamma^{a} k_{a} +s ) }
  { ( k^{2} -s^{2}) ^{3} }  \nonumber       \\ 
            & &- \frac{1}{2} \gamma^{a} J^{cd} R_{cda \mu} k^{ \mu} 
\frac{1}{ ( k^{2} -s^{2}) ^{2} } \;    ,
\end{eqnarray} 
where $J^{ab} = \frac{1}{4} [\gamma^{a},\gamma^{b}]$, and the
Latin and Greek indices refer to a local orthonormal frame and
general coordinate system, respectively. 
Eq. (11) is the linear approximation for
the Green function $G_{F}(k;s)$ in the curvature $R$ \cite{10,41,42}.
According to the well known method developed in \cite{15}, to obtain eq. (11)
one should neglect  any terms involving derivatives higher than
that of the second order in the metric tensor expansion.
Now let us consider the effect of nonzero chemical potential $\mu$ on the
system. As is well - known, the fermion-number density is directly related to 
the chemical potential $\mu$.
Mathematically, the presence of nonzero chemical potential is
realised by shifting the energy levels $k_{0}$ in the propagator
$G_{F}(k;s)$ by the amount of $\mu $ \cite{16}.
Thus, we are in order to evaluate the effective potential $U(
\sigma )$ in Eq. (10) under effects of both $R$ and $\mu$.
Using the contour integration method \cite{16}, we can perform
the integration over momentum $k^{\mu}$.  Denote $I_1$ the integral of the
first term in $G_{F}(k;s)$ over $k$ and $ s$. Its calculation proceeds
as follows: first, the procedure of integration over $k_{0}$, denoted as
$I_{1}^{\prime} $, gives the result:
\begin{eqnarray}
I_{1}^{\prime}(k,s) &\equiv&\mbox{tr} \int \frac{dk_{0}}{2\pi}\frac{\gamma^{0}(k_{0}+\mu)
+\gamma^{i} k_{i}+s }{(k_{0}+\mu)^{2}-E_{k}^{2}   } \nonumber \\
                    &=     & \frac{2}{\pi} \int^{i \infty}_{-i \infty} \frac{s\; dz}{z^{2}-E_{k}^{2}}
  + \frac{2}{\pi} \oint_{C} \frac{s\; dz}{z^{2}-E_{k}^{2}} \nonumber \\   
                    &=     & \frac{2}{\pi} \int^{i \infty}_{-i \infty} \frac{s\; dz}{z^{2}-E_{k}^{2}}
+\frac{2is}{ E_{k}}\theta(\mu -E_{k}).
\end{eqnarray} 
Here, $E_{k}^{2} \equiv k^{2}+s^{2}$, the contour $C$ is given
in Fig. 1, and the unit step function
$\theta(x)=1 \mbox{ for } x>0, \;  \theta(x)=0 \mbox{ for } x<0$
has been used.  Thus we get
\begin{eqnarray}
I_{1} &\equiv&-i \int^{\sigma}_{0}ds \int \frac{d^{2}k}{(2\pi)^{2}}I_{1}^{\prime}(k,s)\nonumber \\
      &=     &\sigma^{2}\left[ \frac{ \sigma}{3 \pi}-\frac{\Lambda}{\pi^{2}}\right]+
\theta(\mu-\sigma) \left[ \frac{\mu}{2 \pi}\sigma^{2}-\frac{1}{3
\pi}\sigma^{3}\right]+\theta(\sigma -\mu) \frac{\mu^{3}}{6},
\end{eqnarray} 
where $\Lambda$ is the cutoff parameter. Here and in the
following discussions, we may confine ourselves to the $\sigma
\geq 0$ region due to a reflection symmetry
$\sigma\leftrightarrow -
\sigma$ of the effective potential $U(\sigma)$. However, note
that this symmetry is broken when the system selects one of the
two ground states.  In a similar way one finds the contributions
$I_{2},I_{3},I_{4}$ of the
remaining terms of $G_{F}(k;s)$ to the potential (10):
\begin{eqnarray}
I_{2}&=& \frac{R}{24 \pi}\left[ -\sigma +\theta(\mu-\sigma)(\sigma-\frac{1}{2\mu}\sigma^{2})+\theta(\sigma-\mu)\frac{\mu}{2} \right], \nonumber \\
I_{3}&=&\frac{1}{12 \pi}\left[ R\; \sigma -R_{00}[\theta(\mu-\sigma) (\sigma-\frac{1}{2\mu}\sigma^{2})+\theta(\sigma-\mu) \frac{\mu}{2}] \right], \nonumber \\
     &=& \frac{1}{12 \pi} R\; \sigma,  \nonumber \\
I_{4}&=&0.
\end{eqnarray}
In the third line of Eq.(14), we have used the relation $R_{00} =0$, which
follows from the metric Eq.(1) of the spacetime under consideration.
However, the fourth line of Eq.(14) is due to a relation 
tr$ [\gamma^{i}\gamma^{j}\gamma^{k}]=0$.
At this stage it is convenient to introduce the mass parameter $M$
instead of the coupling constant $\lambda$ 
by the following way \cite{2}: 
\begin{eqnarray}
\frac{1}{\lambda^{2}}&\equiv& 4 \int^{\Lambda} \frac{d^{3}k_{E} }{(2 \pi)^{3}} \frac{1}{ k_{E}^{2}+ M^{2}} \nonumber \\ 
                  &   =  &\frac{2}{\pi^{2}} \Lambda-\frac{1}{\pi} M.
\end{eqnarray}
So, we shall consider the case $\lambda
>\lambda_{c}$ only, where $\lambda^{-2}_{c}
\equiv 4\int^{\Lambda} d^{3}k_{E}(2\pi)^{-3}k^{-2}_{E}$.
Summing up all terms $I_{i}$ in Eq. (13) and (14) and inserting above 
equation into Eq. (10), one sees that the two
$\Lambda$-dependent terms cancel out, and thus the finite
effective potential to one-loop order is obtained.  Then, the
$\mu$- and $R$-dependent one-loop contributions
$U^{1}_{R\mu}(\sigma)$ to the potential $U(\sigma)$ are
completely separated from the Minkowski-space result:
\begin{equation}
U( \sigma) =U_{F}(\sigma)  +U_{ R\mu }^{1} (\sigma  ) ,
\end{equation}
where $U_{F}(\sigma)$ is the effective potential of the original theory 
in flat Minkowski spacetime. Here
\begin{eqnarray}
U_{F}(\sigma)               &=& \frac{ \sigma^{2}}{3 \pi} \left[ \sigma- \frac{3}{2} M \right]       ,    \nonumber \\   
U_{ R \mu}^{1}(\sigma)    &=& \frac{R}{ 24  \pi} \sigma + \frac{1}{ \pi} \theta 
( \mu- \sigma) \left[ (\mu - \frac{2}{3} \sigma) \frac{ \sigma^{2}}{2} + \frac{R}{24} 
( \sigma -\frac{ \sigma^{2}}{2 \mu} ) \right] \nonumber \\  
                            & & +\frac{1}{  6 \pi} \theta( \sigma -\mu) \left[
\mu (\mu^{2} + \frac{R}{8}) \right].
\end{eqnarray}
In this expression one may find  the following two facts.
Firstly, $ U_{ R \mu}^{1}(\sigma)$ is finite and, as $R,\mu \rightarrow 0$, $  
U_{ R \mu}^{1}(\sigma)$ vanishes.  Thus the renormalisation procedure is 
identical to the case of Minkowski spacetime. 
Secondly, in the limit $\mu, R \rightarrow 0$, 
$U(\sigma)$ is reduced to the Minkowski-space
effective potential $U_{F}(\sigma)$.
It is well established that there are two distinct phases in the
three - dimensional GN model
\cite{1,2,6}.  For a weak coupling phase with the coupling $\lambda
<\lambda_{c}$ we have $\langle \sigma \rangle =0$. Thus the
fermions are massless and the chiral symmetry remains intact.
However, for the strong coupling phase $\lambda > \lambda_
{c}$, $\sigma$ field has nonzero vacuum expectation value
$\langle \sigma \rangle =M$ , so the chiral symmetry Eq. (2) is
dynamically broken and fermions aquire mass, which is equal to
the mass parameter $M$ from Eq. (15).
For the simplicity of our analysis in the next sections, we
shall introduce the following rescaled dimensionless quantities
defined as $\util ( x) \equiv \pi U( \sigma)/ \mu^{3},\; \rtil
\equiv R/\mu^{2}, \; x \equiv \sigma / \mu$, and $  \mbar \equiv \mu/ M.$ 
In terms of these quantities, Eq. (16) is rewritten in the much simpler form:
\begin{equation}
\util (x)= \left\{  \begin{array}{ll}
                  (1-\frac{ 1}{ \mbar}-  \frac{ \rtil}{24} )\frac{x^{2}}{2} +\frac{\rtil  x} 
{12} ,                        &  \mbox{  for  $  x<1 $}    \\
                     (x- \frac{3}{2 \mbar}  )\frac{ x^{2}} {3} +\frac{\rtil x}{ 24} +
\frac{1}{6}( 1+ \frac{\rtil}{8}),   &   \mbox{  for $ x \geq 1$},    
\end{array}
            \right.  
\end{equation}
where one sees that $\util(x)$ is a continuous function at
$x=1$.  We also wish to find the induced fermion mass $\langle
\sigma \rangle$ as a function of curvature $R$ and chemical
potential $\mu$.  Then, the gap equation for the fermion
mass can be obtained by taking the derivative of the effective
potential $\util (x)$ with respect to $x$, and so we obtain
\begin{equation}
0               = \left\{ \begin{array}{ll}
                     (1- \frac{1}{\mbar}-  \frac{\rtil}{24} ) x +\frac{\rtil}{12},                                          &     \mbox{  for  $  x<1 $}    \\
                     x^{2}- \frac{x}{\mbar}   +\frac{\rtil}{24}, 
                                   &   \mbox{  for $ x \geq 1$}.    
                    \end{array}
           \right. 
\end{equation}
\section{  Restoration of chiral symmetry }
Now we shall analyze in detail the effective potential Eq. (18)
in order to investigate
the phase structure of the model in the $(R,\mu)$ plane.
The fermion mass $\langle \sigma  \rangle $ will be derived which depends
on $R$ and $\mu$ and the nature of the phase
transitions will be discussed.  To clarify our discussion, we
shall consider three distinct cases: $\mu \neq  0$ and $R=0$,
then $R \neq 0$ and $\mu =  0$, and finally $R \neq 0$ and $\mu
\neq  0$.
\subsection* {A.  The case $\mu \neq  0$ and $R=0$. }
Let us first examine the effect of nonzero chemical potential on the system.
In the limit $R \rightarrow  0$, the effective potential Eq. (18) is 
reduced to a simple form:
\begin{equation}
\util (x)= \left\{ \begin{array}{ll}
                    (1- \frac{1}{\mbar})\frac{x^{2}}{2},             &   \mbox{  for  $x<1$} \\
                    (x- \frac{3}{2 \mbar})\frac{ x^{2}}{3} +\frac{1}{6},   &   \mbox{  for  $x \geq 1$.}
                   \end{array}
            \right.
\end{equation}
To see a phase transition as $\mbar$ increases from a
broken phase to a symmetric one, it is necessary to examine the behavior
of $\util (x)  $ as a function of $\mbar  $.  It is possible to find the
following two properties of $\util (x)  $.  For $\mbar >1  $,
$\util (x)  $ is a monotonically increasing function of $x$, and
so the global minimum of $\util (x) $ occurs at $x=0$. While
for $ \mbar <1$ $\util (x)$ has a global minimum at $x= 1/
\mbar  $ with the value:
\begin{equation}
\util \left( x=\frac{1}{  \mbar}\right) = -\frac {1}{6}  \frac{( 1- \mbar^{3}) } {\mbar^{3} }.
\end{equation}
These facts indicate that the system undergoes a phase transition from the
$\langle \sigma \rangle=M$ 
state to the  $\langle \sigma \rangle=0$ state at
the critical value $\mu_{c}$ of the chemical potential, given as 
\begin{equation}
\mu_{ c  }= M.
\end{equation}
Solving the gap equation for the induced fermion mass, Eq. (19) with $R=0$,
one can find that 
\begin{equation}
 \langle \sigma \rangle = M
\end{equation}
below $\mu_{c}$, and  $\langle \sigma \rangle= 0$ above
$\mu_{c}$. Except at $\mu = \mu_ {c}$, the order parameter
$\langle \sigma
\rangle$ does not depend on the value of $\mu$.  That is, the value
of order parameter $\langle \sigma \rangle$, which minimizes the
potential, jumps discontinuously from $\sigma=M$ to $
\sigma =0$ at the transition point $\mu_{c}$. 
Hence, at the point $\mu=\mu_c$ we have a first order phase transition 
from a massive chirally broken phase to a massless chirally invariant phase
of the model.
\subsection* {B. The case $R \neq  0$ and $\mu=0$}
In this case only the effect of curvature on the system will be considered.
In the limit $\mu \rightarrow 0$, the general effective potential Eq. (18) 
has the following form:
\begin{eqnarray}
U( \sigma) & = & U_{F}(\sigma) +U_{R}^{1} (\sigma )  \nonumber  \\
           & = & \frac{ \sigma^{2}}{3 \pi} \left( \sigma- \frac{3}{2} M \right) + \frac{R}{ 24  \pi} \sigma . 
\end{eqnarray}
This expression coincides with that obtained in \cite{43}. 
>From Eq. (24) one can see that in the region of small values of $\sigma$ the
dominant contribution to $U( \sigma )$ comes from the
$R$-dependent linear term in $\sigma$.  Thus, there is a
potential barrier between $\sigma =0$ and second local minimum
of $U(\sigma)$. As a result, it turns out that as the curvature $R$ increases
the discontinuous phase
transition  occurs from a chirally broken phase to a symmetric one.
The critical value of the curvature $R_{c}$, at which a first
order phase transition occures, is determined by the following two
conditions:
\begin{equation}
U^{\prime} (\sigma_{0})=0 \mbox{  and  } U (\sigma_{0})=0, 
\end{equation}
where $\sigma_{0}$ denotes second nonzero local minimum of the potential.
Furthermore, one may find that only for $R > R_{c}$ the minimum of the 
potential at the symmetric point $\sigma=0$ is 
lower than asymmetric local minimum at a nonzero $\sigma_0$.
>From the gap equation Eq.  (19) with $\mu=0$ one can evaluate
the local minimum of the potential $\sigma _{0}$,
\begin{equation}
 \sigma_{0} = \frac{M}{2}\left(1+ \sqrt{1- \frac{1}{6} \frac{R}{M^{2}} }\;  \right) ,
\end{equation}
which at the same time equals to the  fermion mass $\langle
\sigma \rangle$, induced under
the influence of curvature $R$ for $R$ $<$ $R_{c}$ only.
Thus, applying the critical condition Eq. (25) to the effective 
potential Eq. (24), one can obtain the critical curvature
\begin{equation}
R_{c}= 4.5  \; M^{2}.
\end{equation}
The phase transition under the influence of $R$ is a first-order one since it 
occurs discontinuously.
\subsection* {C. The case $R \neq  0$ and $\mu \neq 0$}
In given Subsection we are going to explore the general case
when the system is specified by the 
curvature and finite chemical potential.
To investigate the vacuum structure of the system as $R$ and
$\mu$ are varied, one must first examine the behavior of the
potential $\util (x)$ as a function of $R$ and $\mu$.
It is very helpful to sketch qualitatively the effective potential 
$\util (x)$ from Eq. (18). 
For $\mbar >1 \; (\mu >M)$ the global minimum of $\util 
(x)$ occurs only at $x=0$. While for $\mbar < 1$ $(\mu < M)$, the global 
minimum of $\util (x)$ lies at nonzero point certainly.
Therefore, 
when $\mbar <1$, it turns out that the system undergoes a phase
transition from the $\langle \sigma \rangle \neq 0$ vacuum state
to the $\langle \sigma \rangle = 0$ state at certain critical
curvature $\rtil _{c}$ depending on $\mu$ .
Using a much detailed analysis of the effective potential $\util
(x)$ in Eq. (18), one can see that until the system approaches
the critical point with the increase of curvature, the second
local minimum of $\util (x)$ occurs only in the region $ x >1$.
Therefore, in the procedure of determining the critical value of
the curvature $\tilde{R}_{c}$, the effective potential needs to
be considered only in the $x>1$ region in Eq. (18).
 
In this case we can obtain the critical curvature $R_{c}$ 
also using the condition given in Eq. (25), with the only change
$\sigma_{0} \rightarrow x_{0}$, where $x_{0}$ denotes the
second local minimum of the potential. That is, in the present case the phase
transition under investigation  is also a
first-order one.  As can be easily checked from the gap
equation Eq. (19), the second minimum lies at the point
\begin{equation}
x_{0}= \frac{1}{2 \mbar} \left(  1+ \sqrt{1- \frac{ \rtil \mbar^{2}}{ 6} } \; \right) .  
\end{equation}
Thus, the critical condition Eq. (25) with this value for
$x_{0}$ leads to the self-consistent relation on the critical
curvature $\rtil _{c}$:
\begin{equation}
16 x_{0}^{3}- \frac{24 x_{0}^{2}} { \mbar} +(2 x_{0}+1) \rtil _{c} +8=0 ,
\end{equation}
where $x_{0}$ has the value given in Eq. (28), with $\rtil$ replaced by $\rtil _{c}$.
The numerical solutions of Eq. (29) are illustrated in Fig. 2.
Note that as $\mu \rightarrow  0$ the $R_{c}$ approaches $4.5 \; M^{2}$ and
as $R \rightarrow 0$, the $\mu _{c}$ approaches $M$. These
limiting cases have been already discussed in the previous Subsections.
Eq. (28) says that the induced fermion mass $\langle \sigma
\rangle$, with $\langle \sigma \rangle= \mu x_{0}$, does depend
on the curvature $R$ only. That is, $\langle \sigma \rangle$
does not depend on $\mu$, and thus it has the same expression as
Eq. (26).  In Fig. 3, the effective potentials  are given for
four distinct values of $R$ at fixed $\mu = \frac{M}{2}$.
\section {  Summary}
In the present paper we have derived the effective potential of
the three-dimensional Gross-Neveu model in the curved spacetime of the form 
$R^1\times S^2$ and with taking into account the chemical potential
$\mu$ as well.  Then, the critical curvature $R_{c}$ at which
dynamical symmetry breaking disappears has been determined in
terms of the induced fermion mass $M$ in the limit $R,\mu
\rightarrow 0$ and at nonzero chemical potential $\mu$, as given in
Fig. 2. 
As the chemical potential $\mu$ increases, the fermion-number
density increases also.  In Subsections A and C, it has been shown
that the high density influences the symmetry behavior of the
system, and so at the critical value $\mu_{c}$ or at the
corresponding critical number density the chiral phase
transition is occurs.  Then, we have observed 
that the order parameter $\langle \sigma \rangle$ of the phase
transition, corresponding to the minimum of the potential, does not
depend on the value of $\mu$, except at the critical value $\mu=
\mu_{c} $,
even though the phase transition is induced by the chemical potential.
This phenomenon is connected with the fact that the composed
field $\sigma\sim\bar\psi\psi$ is a real field and carries no charge.
It was observed also in two - dimensional GN model 
in $R^1\times S^1$ spacetime \cite{13}, however, in that model
there is another massive phase, in which fermions mass is $\mu$ - dependent
quantity.
In Subsection B, we have shown that $R_{c}=4.5 M^{2}$. This can be
roughly seen from the following two facts. Firstly, on
dimensional grounds the critical curvature $R_{c}$ must be
proportional to the square of some quantity with the dimension
of mass.  Secondly, the effective potential for the composite
$\sigma$ field in Eq. (24) has two parameters $R$ and $M$,
and so the remaining parameter apart from $R$ in this theory is $M$.
Note, that our value for $R_{c}$ is valid only in a weak curvature
limit, and thus its more accurate value can be obtained by
considering higher order corrections to scalar curvature $R$.
However, in such improved schemes, it is expected that the
system still show the same qualitative properties as those found in the
previous Sections, including the occurrence of a first order phase transition.
Finally, one may consider the case of negative curvature since
the present method has the advantage of being applicable to any
metric.  Then, Eq. (26) indicates that under the effect of
negative curvature $R$ the minimum of the potential moves
farther from the origin than without the curvature effect.
Therefore, in this case the symmetry restoring phase transition
does not happen.
 
We hope that the above results may be useful for condensed
matter physics as well, and for astrophisical applications, especially
for the desciption of different phenomena in the core of neutron stars.
\section*{ Acknowledgements}
 We are grateful to Prof. S.D. Odintsov and
Dr. P.A. Saponov for reading the manuscript and some critical
remarks as well as to Prof. V.P. Gusynin for useful comments.
D.K. Kim thanks Prof. K.-S. Soh, Prof. C.K. Kim and Prof. J.H. Yee for
helpful discussions.
%\vspace{0.5cm}
%\noindent
%\vspace{0.5cm} \\

\newpage
\begin{flushleft}
{\bf Figure Captions }
\end{flushleft}
{\bf Fig.1}.  The contour C in the complex $k^{0}$ plane. 
{\bf Fig.2}.  The critical curvature $R_{c}/M^{2}$ as a function of
nonzero chemical potential $\mu/M$. In region B, chiral symmetry is broken
and fermions acquire dynamical masses, while in S, the symmetry is restored by
the curvature effect, and fermions become massless.
 {\bf Fig.3}. The effective potential $\pi U(\sigma)/M^{3}$ as a function
of $\sigma /M$ at the fixed value of $\mu/M= 1/2$. Four interesting cases
of $\bar{R}$, where $\bar{R} \equiv R/M^{2}$, are considered, and the critical
curvature $\bar{R}_{c}$ is then numerically obtained: $\bar{R}_{c}=2.96$.
\end{document}